\documentstyle[11pt]{article}

\let\a=\alpha 
\let\b=\beta 
\let\g=\gamma  
\let\d=\delta 
\let\e=\epsilon
 
\let\h=\eta  
\let\k=\kappa
\let\l=\lambda
\let\m=\mu
\let\n=\nu 
 
\let\r=\rho
\let\s=\sigma 
\let\t=\theta

\let\la=\label

\let\bm=\bibitem

\def\nn{\nonumber} 
\def\bd{\begin{document}} 
\def\ed{\end{document}}
\def\ds{\documentstyle} 
\def\v{\wedge}
\def\ba{\begin{array}}
\def\ea{\end{array}}
\def\bea{\begin{eqnarray}}
\def\eea{\end{eqnarray}}
\def\ft#1#2{{\textstyle{{\scriptstyle #1}\over {\scriptstyle #2}}}}
\def\fft#1#2{{#1 \over #2}}
\def\sst#1{{\scriptscriptstyle #1}}
\def\oneone{\rlap 1\mkern4mu{\rm l}}
\def\[{{[}}
\def\]{{]}}
\def\es{\vspace{10pt}}
\def\oneone{\rlap 1\mkern4mu{\rm l}}
\def\hm{{\hat\mu}} 
\def\hn{{\hat\nu}} 
\def\da{{\dot\alpha}}
\def\4{\m_1\cdots \m_4}
\def\5{\m_1\cdots \m_5}
\def\6{\m_1\cdots \m_6}\def\h6{\hm_1\cdots \hm_6}
\def\Hat#1{\widehat{#1}}
\def\qq{\quad\quad}
\def\uC{{\underline C}}

\newcommand{\ho}[1]{$\, ^{#1}$}
\newcommand{\hoch}[1]{$\, ^{#1}$}
\newcommand{\Str}{\rm Str\, }
\newcommand{\be}{\begin{equation}}
\newcommand{\ee}{\end{equation}}
\newcommand{\p}{\partial}
\newcommand{\tb}{\bar\theta}
\newcommand{\cm}{{\cal M}}
\newcommand{\ha}{{\hat A}}
\newcommand{\hb}{{\hat B}}
\newcommand{\hc}{{\hat C}}
\newcommand{\ham}{{\hat M}}

\begin{document}

\begin{titlepage}

\hfill{CTP TAMU-27/96}

\hfill{IC/96/125}

\hfill{hep-th/9609086}

\hfill{September 1996}

\vspace{30pt}

\centerline {\Large\bf The M-Algebra}

\vspace{30pt}

\centerline{ Ergin Sezgin\footnote{Research supported in part by 
NSF Grant  PHY-9411543}}

\vspace{15pt}

\centerline{\it Center for Theoretical Physics, Texas A\&M University,}
\centerline{\it College Station, Texas 77843, U.S.A.}

\vspace{50pt}

\centerline{ABSTRACT}

\vspace{15pt}

We construct a new extension of the Poincar\'e superalgebra in eleven
dimensions which contains super one-, two- and five-form charges. The
latter two are associated with the supermembrane and the superfivebrane
of M-theory. Using the Maurer-Cartan equations of this algebra, we
construct closed super seven-forms in a number of ways. The pull-back of
the corresponding super six-forms are candidate superfivebrane
Wess-Zumino terms, which are manifestly supersymmetric, and contain
coordinates associated with the new charges.

%\pagebreak

\end{titlepage}

As is well known, supergravity theories in diverse dimensions admit a
variety of super $p$-brane solitons. It is now widely appreciated that these 
objects play an important role in  nonperturbative string physics.  While it is 
difficult to analyze the dynamics of super $p$-branes in general, they do
possess some algebraic properties that are more amenable to study. One of these
properties is the modification of the Poincar\'e superalgebra in presence of
super $p$-brane solitons \cite{p,pkt1}. 

Let us consider the case of eleven dimensional supergravity. It admits
supermembrane soliton \cite{ds} which modifies the 11D Poincar\'e
superalgebra as follows:
\be
\{ Q_\a, Q_\b\} = \g^\m_{\a\b}\ P_\m + \g_{\m\n\a\b}\ Z^{\m\n}\ ,\la{pa}
\ee
where the topological charge $Z^{\m\n}$ arises due to the presence of a
supermembrane configuration in spacetime. The Poincar\'e supercharges 
$Q_A=(P_\m, Q_\a)$ generate supertranslations in flat superspace
parametrized by the coordinates $Z^M=(X^\m,\t^\a)$, $\m=0,1,...,10$, where 
$\t^\a$ are anticommuting 32 component Majorana spinors. The notation
for the Dirac $\g$-matrices is self explanatory. 

Given a supermembrane soliton in $D=11$, the Noether supercharge per
unit membrane area is defined as an integral over an eight dimensional
transverse space-like surface. The Poisson bracket algebra of the supercharges 
yields the result (\ref{pa}). The $p$-form charges arising in this way are
variably referred to as topological charges, or Page charges, or
sometimes as central charges. Strictly speaking, they are not central
because they do not commute with Lorentz generators, except for $p=0$. 

The occurrence of the topological charge $Z^{\m\n}$ in 11D Poincar\'e
superalgebra can also be understood from the supermembrane worldvolume
point of view. As shown in \cite{pkt1}, the presence of a Wess-Zumino
term in the supermembrane action, which is supersymmetric only up to a
total derivative term, modifies the algebra of supercharges precisely as
in (\ref{pa}), with the topological charge realized as
\be
Z^{\m\n}=  \int d^2 \s\ \e^{0ij} \p_i X^\m \p_j X^\n\ , \la{tc}
\ee
where $\s^i\,(i=0,1,2)$ are the worldvolume coordinates. These can also
be viewed as the topological charges associated with the identically
conserved topological current $J_T^{i\m\n}=\e^{ijk} \p_j X^\m \p_k
X^\n$. If at fixed time the membrane defines a nontrivial 2-cycle in
space, then the above integral will be nonzero \cite{pkt1}. 

It is known that $D=11$ supergravity also admits a superfivebrane
soliton \cite{guven}. Although the corresponding superfivebrane action
has yet to be constructed, it is reasonable to expect that five-form
topological charges will be associated with them.

Recently, it has been argued that \cite{bars} the $p$-brane topological
charges discussed above are closely associated with the boundaries of
$p$-branes and as well as the topology of the background geometry in
which they propagate. It is furthermore observed \cite{bars} that more
general kinds of topological charges may emerge in general backgrounds
other than flat spacetime. 

Sometime ago, Bergshoeff and the author \cite{bs1,es2} indeed found an extension
of the 11D Poincar\'e superalgebra with the extra generators $Z^{\m\n}$ and
$Z^{\m\a}$. A Kac-Moody extension of the algebra naturally led to the following
generators:
\be
Z^{\m\n}(\s) = \e^{0ij}\, L_j^\m L_k^\n\ ,\qq
Z^{\m\b}(\s) = \e^{0ij}\, L_j^\m L_k^\b\ ,\qq
Z^{\a\b}(\s) = \e^{0ij}\, L_j^\a L_k^\b\ , \la{loop}
\ee
where $\s$ refers to the membrane worldvolume coordinates, and the
supersymmetric line elements are given by
\be
L_i^\m=\p_i X^\m+\ft12 \g^\m_{\a\b}\ \t^\a\p_i\t^\b\ ,\quad\quad
L_i^\a=\p_i\t^\a\ . \la{li}
\ee

Later, the extended version of 11D Poincar\'e superalgebra found in
\cite{bs1} was generalized to include the generators $Z^{\a\b}$
\cite{bs2}, and a spinor generator $Z^\a$ \cite{es}. As far as we know,
there is no superstring soliton in $D=11$, and therefore the occurrence
of the spinor generator $Z^\a$ is somewhat of a mystery at present (see,
however, \cite{tc,es}). Its occurrence in 10D Poincar\'e superalgebra is
natural, and indeed Green \cite{g} discovered it sometime ago. This
superalgebra was used in an interesting way by Siegel \cite{s1}, who
introduced a new coordinate for the extra fermionic generator, and
reformulated the Green-Schwarz superstring such that the full action,
including the Wess-Zumino term, exhibited manifest supersymmetry.
Siegel's result forms an important part of our motivation for this work,
and therefore we shall come back to this point later.

Interestingly enough, extensions of the 11D Poincar\'e superalgebra
which contains some of the generators mentioned above were considered
long ago, before the discovery of higher super $p$-branes. In
particular, D'Auria and Fr\'e \cite{df1} considered the extra generators
$Z^{\m\n},\, Z^{\m_1\cdots \m_5}$ and $Z^\a$, in their attempts to gauge
the $D=11$ supergravity. The issue of whether the dual formulation of
$D=11$ supergravity with a six-form potential existed also emerged in a
related study which used these extensions \cite{df2}. 

It is clearly of interest to unify and extend the results mentioned
above, in a way that would take into account the existence of the
superfivebrane in $D=11$. Further motivations for studying the general
topological extensions of 11D Poincar\'e superalgebra are: (a) They may
provide a powerful tool, in the framework of an extended super
Poincar\'e geometry with new $p$-form coordinates, for probing
supermembrane-superfivebrane duality, (b) the topologically extended
super Poincar\'e geometry may provide important ingredients for the
construction of the elusive eleven dimensional superfivebrane action and
(c) knowledge of their representations may shed some light some
algebraic aspects of M-theory, including the spectrum of nonperturbative
states \cite{bars}. 

In this paper, we shall give the most general extension of the 11D
Poincar\'e superalgebra motivated by some geometrical considerations
which will be spelled out below, and that contains super one-, two- and
five-form charges
\be
Z^{\m_1\cdots \m_k\a_{k+1}\cdots \a_p}\ , \quad p=1,2,5\ ;\ k=0,1,...,p\
,\la{zs}
\ee
in addition to the usual super Poincar\'e generators $P_\m$ and $Q_\a$.
Our result contains the ones in \cite{bs1,bs2,es,df1} as special cases.
For short, we will refer to this algebra as the M-algebra. We have
in mind, of course, the role it is expected to play in $M$-theory.
Interestingly, we find that the existence of the super five-form charges
in the algebra necessarily requires the presence of the super two-form
charges, while the reverse is not true. Moreover, it turns out that some
of the super two-form charges cease to (anti) commute with each other. 

We expect that there will be a number of interesting properties of the
M-algebra which will be uncovered in the future. For the purposes of
this note, however, it will suffice to (a) present the algebra, and (b)
to construct closed super seven-forms that live on the full supergroup
manifold. At the end of the paper, we shall comment on their possible
use in the construction of candidate Wess-Zumino terms for the eleven
dimensional superfivebrane.

We now turn to the description of the M-algebra. Let us denote the
generators of the algebra collectively as $T_\ha $. We consider the
generators
\be
T_\ha = (Q_A,~Z^A,~Z^{AB},~Z^{A_1\cdots A_5})\ , \quad\quad 
                              Q_A = (P_\m, Q_\a)\ ,\la{gen}
\ee
where the $Z$-generators are totally graded antisymmetric. The
M-algebra can be written as
\be
\[ T_\ha, T_\hb\} = f_{\ha\hb}{}^\hc\ T_\hc\ , \la{sc}
\ee
where the structure constants will be given shortly. In the dual basis,
one defines the Maurer-Cartan super one-forms
\be
e^\ha = dZ^\ham\ L_\ham{}^\ha\ , \la{mcf}
\ee
where $dZ^\ham$ are the differentials on the supergroup manifold based on the
M-algebra. We can also define the supersymmetric line elements 
\be
L_i^A=\p_i Z^\ham L_\ham{}^\ha\ .  \la{ssle}
\ee
Explicit expressions for these objects can be obtained straightforwardly from
\be
U^{-1}\p_i U = L_i^{\ha}\ T_{\ha}\ ,\la{udu}
\ee
where $U$ is a group element, which can be parametrized in terms of the  usual 
superspace coordinates, and the new $\phi$-coordinates associated with the
$Z$-generators as follows:
\be
U=e^{  \phi_{\5} Z^{\5} } \cdots e^{ \phi_{\a_1\cdots \a_5} Z^{\a_1\cdots
\a_5} }\ 
e^{ \phi_{\m\n} Z^{\m\n} }\cdots e^{ \phi_{\a\b} Z^{\a\b} }\ 
e^{ \phi_\m Z^\m }\ e^{ \phi_\a Z^\a }\ e^{X^\m P_\m}\ e^{\t^\a Q_\a}\ . \la{uu}
\ee
The details of $L_i^{\ha}$, as well as the M-group transformations
under which they are invariant, are not particularly illuminating.
However, the interested reader can find useful formulae for their
computation, as well as explicit expressions for $L_i^{\ha}$ in the case
of $Q_A$ and $Z^{AB}$ generators in \cite{bs2}. 

As is well known, the Maurer-Cartan structure equations 
\footnote{Our conventions for super $p$-forms are those of \cite{bst}.
In particular, the components of a super $p$-form $F$ are defined by
$F=\ft1{p!}\ e^{M_1}\v\cdots e^{M_p}\ F_{M_p\cdots M_1}$, and the
exterior derivative by $dF=\ft1{p!}\ e^{M_1}\v\cdots e^{M_p}\v e^M\ \p_M
F_{M_p\cdots M_1}$. Furthermore, given a super $q$-form $G$, we have the
rule: $d(F\v G)=F\v dG+(-1)^q\ dF\v G$.}
\be
de^\ha = -\ft12\ e^\hb\v e^\hc\ f_{\hc\hb}{}^\ha\ , \la{mce}
\ee
contain equivalent information about the algebra. The fact that the
Jacobi identities are satisfied is, of course, encoded in the
integrability condition $d^2 e^\ha =0$. It is convenient to present our
results first in the form of Maurer-Cartan equations. The strategy we
have followed to determine these equations is very simple: We have
parametrized the algebra in the most general possible way that contains
the components of the following super forms as structure constants: 
\bea
T^\m &=& -\ft12 e^\a\v e^\b\ \g^\m_{\a\b}\ , \la{tf}\\
H^{(0)}_3 &=& e^\m\v e^\a\v e^\b\ \g_{\m\a\b}\ , \la{h3f}\\
H_4 &=& \ft14 e^\m\v e^\n\v e^\a\v e^\b\ \g_{\m\n\a\b}\ , \la{h4f}\\
H^{(0)}_7 &=& \ft1{5!} e^{\m_1}\v \cdots \v e^{\m_5}\v e^\a\v e^\b\ 
              \g_{\m_1\cdots \m_5\a\b}\ . \la{h7f}
\eea
Noting that the structure equations for $e^\m$ and $e^\a$ will not be modified
compared to those in ordinary Poincar\'e superspace, one can verify that the
four-form $H_4$ is closed, thanks to the following well known identity 
\be
 \g_{\m\n(\a\b}\,\g^\n_{\g\d)} = 0\ , \la{id1}
\ee
which holds in $D=4,5,7,11$ \cite{ac}. The super-forms $H^{(0)}_3$ and
$H^{(0)}_7$ are not closed. However, as we will see later, they can be
modified so as to be closed in the full M-algebra.

The occurrence of (\ref{tf}) and (\ref{h4f}) can be understood from the
structure of the known super $p$-brane actions \cite{bs1,es2,ploop}. The
inclusion of (\ref{h7f}) is motivated by superfivebrane considerations,
and the assumption that there may exist a dual formulation of 11D
supergravity in which, both, the three-form and six-form potentials occur.
Finally, we have included (\ref{h3f}) for the sake of completeness; a point 
which will become more transparent below. 

The parametrization of the ansatze in a manner described above contains
a large number of parameters. We have determined these parameters by
explicit computation of all the integrability conditions $d^2 e^\ha =0$.
These conditions not only are sufficient to determine all the parameters
in our ansatze, but they also provide several cross checks, since we
obtain an overdetermined system of equations for these parameters. It is
also worth mentioning that, in addition to repeated use of (\ref{id1}),
we have also used the following identity \cite{df2}
\be
\g_{\l(\a\b} \g^{\l\m\n\r\s}_{\g\d)}-
        3 \g^{[\m\n}_{(\a\b} \g^{\r\s]}_{\g\d)}=0\ .  \la{id2}
\ee
Also useful are the following identities
\bea
&&\g_{\m(\a\b} \g^\m_{\g\d)}+\ft1{10}\g_{\m\n(\a\b}
      \g^{\mu\nu}_{\g\d)}=0\ , \la{id3}\\
&&\g_{\m(\a\b} \g^\m_{\g\d)}+ 
       \ft1{720}\g_{\m_1\cdots \m_5 (\a\b}
            \g^{\m_1\cdots \m_5}_{\g\d)}=0\ . \la{id4}
\eea
The identities (\ref{id2}) and (\ref{id3}) follow straightforwardly from
the main identity (\ref{id1}), and the identity (\ref{id4}) can be
easily derived from (\ref{id2}). 

With the preliminaries thus explained, we are now ready to present our
results. We propose the following set of Maurer-Cartan equations:
\bea
de^\m &=& -\ft12~e^\a\v e^\b\ \g^\m_{\a\b}\ , \nn\\
\es
de^\a &=& 0\ , \nn\\
\es
de'_\m &=& -\ft12 e^\a\v e^\b\ \g_{\m\a\b}\ , \nn\\
\es
de'_\a &=& -e^\b\v e^\m\ \g_{\m\a\b}+(1-\l-\tau)\ e^\b\v e'_\m\ \g^\m_{\a\b}
         -\ft{\l}{10}\ e^\b\v e_{\m\n}\ \g^{\m\n}_{\a\b}\ \nn\\
      &&  -\ft{\tau}{720}\ e^\b\v e_{\m_1\cdots \m_5}\ 
           \g^{\m_1\cdots \m_5}_{\a\b}\ ,\nn\\
\es
de_{\m\n} &=& -\ft12 e^\a\v e^\b\ \g_{\m\n\a\b}\ ,\nn\\
\es
de_{\m\a} &=& -e^\b\v e^\n\ \g_{\m\n\a\b}- 
              e^\b\v e_{\m\n}\ \g^\n_{\a\b}\ , \nn\\
\es
de_{\a\b} &=& \ft12 e^\m\v e^\n\ \g_{\m\n\a\b} -\ft12 e_{\m\n}\v e^\m\
\g^\n_{\a\b}
     			     -\ft14 e_{\m\g}\v e^\g\ \g^\m_{\a\b}
             -2 e_{\m\a}\v e^\g\ \g_{\b\g}^\m\ ,\nn\\
\es
de_{\5} &=& -\ft12 e^\a\v e^\b\ \g_{\5\a\b}\ , \nn\\
\es
de_{\4\a} &=& e^\b\v e^\tau\ \g_{\tau\4\a\b}+ e^\b\v e_{\tau\4}\ \g^\tau_{\a\b}
               -6 e^\b\v e_{\m_1\m_2}\ \g_{\m_3\m_4\a\b}\ ,\nn\\
\es
de_{\m\n\r\a\b} &=& \ft12 e^\s\v e^\tau\ \g_{\s\tau\m\n\r\a\b} 
                    -\ft12 e_{\m\n\r\s\tau}\v e^\s\ \g^\tau_{\a\b}
                    +\ft14 e_{\m\n\r\tau\g}\v e^\g\ \g^\tau_{\a\b}   \nn\\
																&& -2 e_{\tau\m\n\r\a}\v e^\g\ \g^\tau_{\b\g}
                   +\ft32 e^\tau\v e_{\tau\m}\ \g_{\n\r\a\b}
                   -3 e^\tau\v e_{\m\n}\ \g_{\r\tau\a\b} \nn\\
                && -\ft34 e^\g\v e_{\m\g}\ \g_{\n\r\a\b}
                   -6 e^\g\v e_{\m\a}\ \g_{\n\r\b\g}
                   +\ft32 e_{\tau\m}\v e_{\n\r}\ \g^\tau_{\a\b}\ , \nn\\
\es
de_{\m\n\a\b\g} &=& 2 e^\s\v e_{\s\tau\m\n\a}\ \g^\tau_{\b\g}
                    + e^\d\v e_{\tau\m\n\d\a}\ \g^\tau_{\b\g}  
                    +5 e^\d\v e_{\tau\m\n\a\b}\ \g^\tau_{\g\d} \nn\\
                && -10 e^\tau\v e_{\m\a}\ \g_{\n\tau\b\g}
                    -5 e^\d\v e_{\a\b}\ \g_{\m\n\g\d}
                    -2 e^\tau\v e_{\tau\a}\ \g_{\m\n\b\g} \nn\\
                &&  - e^\d\v e_{\d\a}\ \g_{\m\n\b\g}
                    -6 e_{\tau\m}\v e_{\n\a}\ \g^\tau_{\b\g}
                    -2 e_{\tau\a}\v e_{\m\n}\ \g^\tau_{\b\g}\ ,\nn\\
\es
de_{\m\a_1\cdots\a_4} &=& -e^\n\v e_{\m\n\tau\a_1\a_2}\ \g^\tau_{\a_3\a_4}
               -\ft3{10} e^\g\v e_{\m\n\g\a_1\a_2}\ \g^\n_{\a_3\a_4}
               -\ft65 e^\g\v e_{\m\n\a_1\a_2\a_3}\ \g^\n_{\a_4\g} \nn\\
           && +3 e^\n\v e_{\a_1\a_2}\ \g_{\m\n\a_3\a_4}
              +2 e_{\m\n}\v e_{\a_1\a_2}\ \g^\n_{\a_3\a_4}
              +3 e_{\n\a_1}\v e_{\m\a_2}\ \g^\n_{\a_3\a_4}\ ,\nn\\
\es
de_{\a_1\cdots\a_5} &=& -\ft35 e^\m\v e_{\m\n\a_1\a_2\a_3}\
                              \g^\n_{\a_4\a_5}
              + e^\g\v e_{\m\g\a_1\a_2\a_3}\ \g^\m_{\a_4\a_5}
														+\ft72 e^\g\v e_{\m\a_1\cdots\a_4}\ \g^\m_{\a_5\g} \nn\\
          &&  +6 e_{\m\a_1}\v e_{\a_2\a_3}\ \g^\m_{\a_4\a_5}\ , \la{malg}
\eea
where it is understood that the obvious symmetries of indices on the
left hand side are to be implemented on the right hand side, with unit
strength (anti) symmetrizations. The parameters $\l$ and $\tau$ are
arbitrary. Of course, by rescaling various one-forms, one can introduce
a number of new parameters. However, the consistency of any contraction
has to be checked carefully. We shall come back to this point shortly.

The world indices can be raised and lowered with the 11D Minkowski
metric, and the fermionic indices with the charge--conjugation matrix,
as usual. However, we have found it convenient not to do so in our
calculations, and to always keep the world and spinor indices in a fixed
position (see comment (iii) below). Nevertheless, to avoid any confusion
between the one-forms associated with the supertranslation generators
$Q_A$, and the topological charge $Z^A$, we have used a prime to
distinguish the latter from the former.

To see the structure of the algebra that underlies the Maurer-Cartan
equations (\ref{malg}) more explicitly, it is convenient to go over to
the dual basis. This is easily done by using (\ref{sc}) and (\ref{mce}),
and we thus find the following (anti) commutation rules:
\bea
\{Q_\a, Q_\b\} &=& \g^\m_{\a\b}\ P_\m + \g_{\m\a\b}\ Z^\m 
                  +\g_{\m\n\a\b}\ Z^{\m\n}+\g_{\5\a\b}\ Z^{\5}\ , \nn\\
\es
\[P_\m,Q_\a\] &=& \g_{\m\a\b}\ Z^\b -\g_{\m\n\a\b}\ Z^{\n\b}
              -\g_{\m\n_1\cdots \n_4\a\b}\ Z^{\n_1\cdots \n_4\b}\ , \nn\\
\es
\[P_\m, P_\n \] &=& \g_{\m\n\a\b}\ Z^{\a\b} 
                    + \g_{\m\n\m_1\cdots\m_3\a\b}\ Z^{\m_1\cdots\m_3\a\b}\
,\nn\\
\es
\[Q_\a, Z^\m\] &=& (1-\l-\tau)~\g^\m_{\a\b}\ Z^\b\ , \nn\\
\es
\[P_\l, Z^{\m\n}\] &=& \ft12 \d_\l^\m \left( \g^\n_{\a\b}\ Z^{\a\b}
                        -3\g_{\r\s\a\b}\ Z^{\n\r\s\a\b}\right)
                        +3\g_{\l\r\a\b}\ Z^{\m\n\r\a\b} \ , \nn\\
\es
\[Q_\a, Z^{\m\n}\] &=& -\ft{\l}{10} \g^{\m\n}_{\a\b}\ Z^\b
                        +\g^\m_{\a\b}\ Z^{\n\b} 
                        -6 \g_{\r\s\a\b}\ Z^{\m\n\r\s\b}\ ,\nn\\ 
\es
\[P_\m,Z^{\n\a}\] &=& -2\d_\m^\n\ \g_{\l\tau\b\g}\ Z^{\l\tau\a\b\g}
                     +10\g_{\m\tau\b\g}\ Z^{\n\tau\a\b\g}\ , \nn\\
\es
\{Q_\a, Z^{\m\b}\} &=& \ft14 \d_\a^\b\ \g^\m_{\g\d}\ Z^{\g\d}
                       +2 \g^\m_{\a\g}\ Z^{\b\g}
                   +\ft34 \d_\a^\b\ \g_{\n\r\g\d}\ Z^{\m\n\r\g\d}
                 +6 \g_{\n\r\a\g}\ Z^{\m\n\r\b\g}\ , \nn\\
\es
\[Q_\g, Z^{\a\b}\] &=& -5 \g_{\m\n\g\d}\ Z^{\m\n\a\b\d}
                      - \d_\g^\a\ \g_{\m\n\d\e}\ Z^{\m\n\b\d\e}\ ,\nn\\
\es
\[P_\m, Z^{\a\b}\] &=& -3\g_{\m\n\g\d}\ Z^{\n\a\b\g\d}\ , \nn\\
\es
\[Z^{\m\n}, Z^{\r\s}\] &=& -3 \g^\m_{\a\b}\ Z^{\n\r\s\a\b}\ , \nn\\
\es
\[Z^{\m\n}, Z^{\r\a}\] &=& -6\g^\m_{\b\g}\ Z^{\n\r\a\b\g}
                            +2\g^\r_{\a\b}\ Z^{\m\n\a\b\g}\ , \nn\\
\es
\[Z^{\m\n}, Z^{\a\b}\] &=& -2\g^\m_{\g\d}\ Z^{\n\a\b\g\d}\ , \nn\\
\es
\{Z^{\m\a}, Z^{\n\b}\} &=& -3\g^\m_{\g\d}\ Z^{\n\a\b\g\d}\ , \nn\\
\es
\[Z^{\m\a}, Z^{\b\g}\] &=& 6\g^\m_{\d\e}\ Z^{\a\b\g\d\e}\ ,\nn\\
\es
\[Q_\a, Z^{\5}\] &=& -\ft{\tau}{720} \g^{\5}_{\a\b}\ Z^\b 
                     +\g^{\m_5}_{\a\b}\ Z^{\4\b}\ ,\nn\\
\es
\[P_\l, Z^{\5}\] &=& \ft12 \d_\l^{\m_1}\ \g^{\m_2}_{\a\b}\ Z^{\m_3\cdots
\m_5\a\b}
                      \ , \nn\\
\es
\{Q_\a, Z^{\b\m_1\cdots\m_4}\} &=& \ft14 \d_\a^\b\ \g^{\m_1}_{\g\d}\ 
        Z^{\m_2\cdots\m_4\g\d}+2\g^{\m_1}_{\a\d}\ Z^{\m_2\cdots\m_4\b\g}\ ,\nn\\
\es
\[P_\l, Z^{\b\4}\] &=& 2\d_\l^{\m_1}\ \g^{\m_2}_{\g\d}\ Z^{\m_2\m_3\b\g\d}\
,\nn\\
\es
\[Q_\a, Z^{\m\n\r\b\g}\] &=& \d_\a^\b\ \g^\m_{\d\e}\ Z^{\n\r\g\d\e}
                             +5\g^\m_{\a\e}\ Z^{\n\r\b\g\e}\ ,\nn\\
\es
\[P_\l, Z^{\m\n\r\b\g}\] &=& -\d_\l^\m\ \g^\n_{\d\e}\ Z^{\r\b\g\d\e}\ ,\nn\\
\es
\{Q_\d, Z^{\m\n\a\b\g}\} &=& -\ft3{10} \d_\d^\a\ \g^\m_{\e\k}\ Z^{\n\b\g\e\k}
                              -\ft65 \g^\m_{\d\e}\ Z^{\n\a\b\g\e}\ ,\nn\\
\es
\[P_\l, Z^{\m\n\a\b\g}\] &=& -\ft35 \d^\m_\l\ \g^\n_{\d\e}\ Z^{\a\b\g\d\e}\
,\nn\\
\es
\[Q_\b, Z^{\m\a_1\cdots\a_4}\] &=& \d^{\a_1}_\b\ \g^\m_{\g\d}\
Z^{\a_2\cdots\a_4\g\d}+\ft72 \g^\m_{\b\g}\ Z^{\a_1\cdots\a_4\g}\ .\la{ms}
\eea
Several comments are in order:

(i) The existence of the algebra (\ref{ms}) is highly nontrivial. To
show that the Jacobi identities are satisfied, one makes crucial use of
the $\g$-matrix identity (\ref{id1}) and its consequences
(\ref{id2})-(\ref{id4}). As is well known, the identity (\ref{id1})
holds in $D=4,5,7,11$ \cite{ac}, i.e. precisely the dimensions in which
the supermembrane action of \cite{bst} exists.

(ii) While in the absence of the super five-form generators all the
remaining $Z$-charges (anti) commute with each other, this ceases to be
the case once the super five-form generators are introduced. This is a
surprising feature, since one normally thinks of the topological charges
as coming from antisymmetric products of $L_i^\m$ and $L_i^\a$, which
are expected to have vanishing Poisson brackets with each other. 

(iii) The form of the algebra is suggestive of a geometrization in which
one works with the generators $Q_A$, $Z^A$, $Z^{AB}$, $Z^{A_1\cdots
A_5}$ and use the super torsion tensor $T_{AB}{}^C$, and the graded
antisymmetric tensors $H_{ABC}$, $H_{ABCD}$ and $H_{A_1\cdots A_7}$
defined in (\ref{tf})-(\ref{h7f}) as structure constants. This would
correspond to a rigid version of a curved superspace algebra.
Surprisingly, this does not work, as it can already be
established at the level of the subalgebra containing only the $Z^{AB}$ as
the new generators. This may suggest the existence of an improved version of
the algebra which can be geometrized. Whether this is indeed possible
remains to be seen.

(iv) The first line of this algebra can be put into a $10+2$ dimensional form
\be
\{Q_\a, Q_\b\} = \g_{\hm\hn\a\b}\ Z^{\hm\hn}+\g_{\h6\a\b}\ Z^{\h6}_+\ , \la{12d}
\ee
where $\hm=0,1,...,10,12$, $Q_\a$ is a $32$ component Majorana-Weyl
spinor, and the $66$ component $Z^{\hm\hn}$, together with the self-dual
$462$ component $Z^{\h6}_+$ add up to $528$ generators. However, it is far
from obvious if the full algebra presented above can be casted into a
$10+2$ dimensional form.

(v) While the presence of the super two- and five-form generators are
related to the existence of the supermembranes and superfivebranes of
eleven dimensional supergravity, the occurrence of the super one-form
generator $Z^A$ is somewhat unexpected, and it is a surprising feature
of the above algebra. See, however, \cite{tc,es} where the issue of
superstring in $D=11$ is discussed. In particular, let us note the
existence of the following closed super three-form
\be
H_3= e^\m\v e^\a\v e^\b\ \g_{\m\a\b}
+e^\a\v e^\b \left( (\l+\tau-1)\, e'_\m\ \g^\m_{\a\b}+\ft{\l}{10}\ e_{\m\n}
 \g^{\m\n}_{\a\b} +\ft{\tau}{720}\, e_{\5}\ \g^{\5}_{\a\b}\right)\ .\la{h3n}
\ee
Indeed $dH_3=0$, and expressing $H_3=dC_2$, and using the Maurer-Cartan
equations (\ref{malg}), one finds that
\be
C_2= -e^\a\v e'_\a \ .  \la{b2}
\ee
One might envisage using the pull-back of this super two-form in
constructing a Wess-Zumino term for a superstring action in $D=11$.
Interestingly enough, this form turns out to play a role in the construction
of a novel Wess-Zumino term for superfivebrane, as we shall show
later. However, these constructions raise a number of questions, among
which is the interpretation of the new coordinates involved in the action.

(vi) Once the five-form generator $Z^{\5}$ is included in the algebra,
it is clear that one has to also include the two-form generator
$Z^{\m\n}$, as can be seen from the $\{Q,\{Q,Q\}\}$ Jacobi identity and
the $\g$-matrix identity (\ref{id2}). The reverse is not true, i.e. one
can have the two-form generator without having to introduce the one-
and/or five-form generators, in view of the $\g$-matrix identity
(\ref{id1}). In fact, the super one- and/or five-form generators can be
contracted away consistently. 

Note in particular that the generators $Z^A$ decouple from the algebra
if we set $\l=\tau=0$ and redefine the translation generator as
$P_\m+Z_\m \equiv P'_\m$. While it may be thought that $Z^\m$ can always
be redefined away, there are some global subtleties in doing so, and at
least in the case of 10D superstrings, they have an interesting role to
play in the description of the string winding states \cite{pkt2}. 

(vii) The fact that the super five-form generator requires the presence
of the super two-form generator is related to the fact that a dual
formulation of $D=11$ supergravity containing only the six-form
potential is not possible \cite{vn,df2}. The coexistence of the super
two- and five-form generators in the M-algebra on the other hand,
suggests a formulation of $D=11$ supergravity theory in which both the
three-form and the six-form potentials are used. However, a duality
relation has to be imposed on the relevant field strengths, which then
leads to non-localities \cite{cl}.

(viii) Contracting away the super five-form generators yields the
algebra of \cite{es}. Contracting away the super one-form generator as
well, one obtains the algebra of \cite{bs2}. Setting equal to zero
$Z^{\a\b}$ in addition gives the result of \cite{bs1}. Keeping only the
generators $Z^\a, Z^{\m\n}$ and $Z^{\5}$ gives the algebra studied in
\cite{df1}. 

(ix) The fermionic generators $Z^\a$ and $Z^{\a_1\cdots \a_5}$ commute with all
the other generators, except Lorentz generators.    

(x) Dimensional reduction of the algebra (\ref{ms}) to ten and lower
dimensions is expected to produce similar algebras for super $p$-branes
existing in those dimensions. Aspects of these reductions will be
treated elsewhere. It should be noted, however, that the Type IIB Poincar\'e
superalgebra in 10D, as well as its M-algebra extension, if any,
cannot be obtained in this way.

We now turn to the issue of Wess-Zumino terms based on the algebra (\ref{ms}).
First of all we observe that the super four-form (\ref{h4f}) is closed
within the
full algebra, and writing  $H_4=dC_3$, we find \cite{bs2}
\be
C_3=-\ft16 e^\m\v e^\n\v e_{\m\n}-\ft3{20} e^\m\v e^\a\v e_{\m\a} 
     +\ft1{30} e^\a\v e^\b\v e_{\a\b} \ . \la{b3n}
\ee
As shown in \cite{bs2}, taking the standard $D=11$ supermembrane action
of \cite{bst}, but using the pull-back of this $C_3$ as Wess-Zumino
term, one finds an alternative formulation of the supermembrane, that 
generalizes a similar construction for the superstring due to Siegel 
\cite{s1}. In doing so, one introduces coordinates for all the
generators, including $\phi^{AB}$, but the dependence on the new
coordinates comes as a total derivative term. This is due to the fact
that the exterior derivative of $C_3$ defined in \cite{bst} and that of $C_3$
defined above give the same result, namely the super-four form
(\ref{h4f}), which in turn has no components along the new directions.
Computing the Noether symmetry algebra corresponding to the full left
group action, one directly finds the full algebra of these generators,
without the occurrence of boundary terms in the Noether current that
arise in a formulation with the usual non-manifestly supersymmetric
version of the Wess-Zumino term. 

To construct a Wess-Zumino term for the superfivebrane in eleven
dimensions, we need a super seven-form. The obvious guess would be
(\ref{h7f}), but that form is not closed, as mentioned earlier. The
$\g$-matrix identity (\ref{id2}) suggests the way to modify (\ref{h7f})
to obtain a closed super seven-form as follows:
\be
H_7 =\ft1{5!}\ e^{\m_1}\v \cdots e^{\m_5}\v e^\a\v e^\b\ \g_{\m_1\cdots
\m_5\a\b}
      + H_4\v C_3\ ,\la{h7n}
\ee
Using the equations (\ref{malg}) and the identity (\ref{id2}), we see that
indeed
$dH_7=0$. Furthermore, writing $H_7=dC_6$, we find that $C_6$ is  given  by
\bea
C_6 =&& \ft1{5!\times 77}\, \big( -\ft{77}3\, e^{\m_1}\v \cdots \v e^{\m_5}\v
                                              e_{\m_1\cdots \m_5} 
        +\ft{281}6\, e^{\m_1}\v \cdots \v e^{\m_4}\v e^\a\v 
                     e_{\m_1\cdots \m_4\a}\nn\\
   && +\ft{104}3\, e^\m\v e^\n\v e^\r\v e^\a\v e^\b\v e_{\m\n\r\a\b}         
      -\ft{47}6\, e^\m\v e^\n\v e^\a\v e^\b\v e^\g\v e_{\m\n\a\b\g}\nn\\
   && +5\, e^\m\v e^{\a_1}\v \cdots \v e^{\a_4}\v e_{\m\a_1\cdots \a_4}
      - \ft59\, e^{\a_1}\v \cdots \v e^{\a_5}\v e_{\a_1\cdots \a_5}\nn\\
   && -\ft{131}3\, e^\m\v e^\n\v e^\r\v e^\a\v e_{\m\n}\v e_{\r\a}
        +\ft{50}3\, e^\m\v e^\n\v e^\a\v e^\b\v e_{\m\n}\v e_{\a\b}\nn\\
   && +\ft{20}3\, e^\m\v e^\a\v e^\b\v e^\g\v e_{\m\a}\v e_{\b\g}\ \big)
    \ .\la{b6}
\eea

Interestingly, the super-form (\ref{h7n}) was considered long ago
\cite{df2} within the framework of a subalgebra of (\ref{ms}) in which
only $Z^\a$, $Z^{\m\n}$ and $Z^{\5}$ are kept in addition to the super
Poincar\'e generators. 

The same super-form was also considered in \cite{cl} within the
framework of the usual 11D Poincar\'e superalgebra. It was shown in
\cite{cl} that defining in curved superspace 
\bea
H_4 &=& \ft14 e^a\v e^b\v e^\a\v e^\b\ \g_{ab\a\b} 
        + e^{a_1}\v \cdots \v e^{a_4}\ H_{a_1\cdots a_4}\ ,\la{h4cl}\\
H_7^{(0)} &=& \ft1{5!} e^{a_1}\v \cdots \v e^{a_5}\v e^\a\v e^\b\ 
          \g_{a_1\cdots a_5\a\b} + e^{a_1}\v \cdots \v 
         e^{a_7}\ \e_{a_1\cdots a_7c_1\cdots c_4}\  H^{c_1\cdots c_4}
         \ ,\la{h7cl}
\eea
one finds, via the Bianchi identities $dH_4=0$ and $dH_7^{(0)}=H_4 \v
H_4$, the correct equation of motion for $D=11$ supergravity. However,
as was emphasized in \cite{cl}, if one wishes to work with a super-six
form potential $C_6$ alone, then one has a non-local relationship between
$H_7^{(0)}$ and $C_6$. 

Turning to the super six-form (\ref{b6}) which is defined in the
M-extended Poincar\'e superspace, we can write down a Wess-Zumino
term for a superfivebrane as follows
\be
I_{WZ} = \int {\uC}_6\ , \la{wz1}
\ee
where ${\uC}_6$ is the pull-back of $C_6$. We use a notation in which
the underlining of a target space form indicates its pull-back. This
action is manifestly invariant under the M-group transformations,
including supersymmetry. However, $H_7=dC_6$ equals an expression that
contains $C_3$ as shown in (\ref{h7n}), which in turn has nonvanishing
components in the $e^{AB}$ directions, as shown in (\ref{b3n}).
Therefore, the Wess-Zumino action (\ref{wz1}) contains the coordinates
$\phi_{AB}$ associated with the generators $Z^{AB}$ such that they are
not confined to a total derivative term. This is in contrast to the
supermembrane case where all the dependence on the new coordinates is
contained in a total derivative term \cite{bs2}. 

Using the super three-form $C_3$ of the standard $D=11$ supermembrane
action \cite{bst} in the definition of $H_7$, on the other hand, would
yield a closed super seven-form that strictly lives in the usual
Poincar\'e superspace. However, the resulting Wess-Zumino term would not
be manifestly supersymmetric. 

In the context of usual Poincar\'e superspace, let us focus our
attention to the case of purely bosonic target space background, and
consider the two-form gauge transformations $\d C_3=d \l_2$. From
$H_7=H_7^{(0)}+H_4\v C_3$, noting that $H_7^{(0)}$ is invariant, one
sees that $C_6$ must transform as $\d C_6 = C_3\v d\l_2$ \cite{cl}. Of
course, $I_{WZ}$ given in (\ref{wz1}) is not invariant under these
transformations. However, it has been observed that \cite{ah}, since the
worldvolume fields of the superfivebrane include a fundamental two-form
$B_2$, there is a way to write down a manifestly tensor-gauge invariant
Wess-Zumino term, namely $I_{WZ}=\int( \uC_6 + dB_2\v \uC_3)$, with
$B_2$ transforming as $\d B_2=\l_2$. 

Turning to the case of superfivebrane in the context of the M-extended
Poincar\'e superspace, we can consider, in analogy with the case discussed
above, the following Wess-Zumino term:
\be
I_{WZ}'= \int \left( {\uC}_6 + dB_2\v \uC_3\right) \ , \la{wz2}
\ee
where, we recall that $C_6$ and $C_3$ are given in (\ref{b6}) and
(\ref{b3n}). This term, just as in the case of (\ref{wz1}) discussed
earlier, contains coordinates other than $X^\m$ and $\t^\a$, which are
not contained in a total derivative term. The target space is flat or
curved M-extended superspace. In the latter case relations similar
(\ref{h4cl}) and (\ref{h7cl}) can be utilized. However, since
super-form coordinates occur in the action, it is not altogether clear
what this implies for $D=11$ supergravity, and whether it can lead its
dual formulation in a novel way.

Going back to the issue of Wess-Zumino terms, it should be emphasized
that one can construct a number of distinct and manifestly
supersymmetric Wess-Zumino terms, within the context of the M-algebra,
by taking various Lorentz-invariant combinations of the left-invariant
super-one forms $L^\ha $. However, presumably not all of these terms are
relevant for the seeked superfivebrane action with the right properties.

To illustrate the point about the variety of ways in which a
superfivebrane Wess-Zumino term can be constructed in our framework, we
present an example which is particularly interesting because it makes
use of the stringy coordinates $Z^A$. Consider the super seven-form
\be
H''_7 = H_4\v H_3\ , \la{h7nn}
\ee
where $H_4$ is defined in (\ref{h4f}) and $H_3$ in (\ref{h3n}). Since
$H_4$ and $H_3$ are closed, so is $H_7$. Moreover, writing $H'_7=dC'_6$,
we have
\be
C''_6= C_3\v H_3\ ,\la{b6nn}
\ee
(up to an irrelevant closed form) where $C_3$ is given in (\ref{b3n}) and
$H_3$ in (\ref{h3n}). Unlike in the case of $H_7$ which can be
formulated in ordinary Poincar\'e superspace, the existence of $H''_7$
requires the M-extended superspace based on the M-algebra, or a
suitable subalgebra thereof. Using (\ref{b6nn}), we can construct a third
kind of Wess-Zumino term given by
\be
I''_{WZ}= \int \uC''_6\ . \la{wz3}
\ee
This is manifestly supersymmetric and tensor-gauge invariant. However,
whether it can be used in the construction of a sensible superfivebrane
action, and if so, exactly which $D=11$ supersymmetric field theory it
may describe, remains to be seen.

So far, we have discussed the superfivebrane Wess-Zumino terms. As far
as the kinetic term is concerned, matters are somewhat more complicated.
Even in the case of a minimal target superspace without any new
coordinates, the full kinetic term is not known. For the progress made
in this front, and a discussion of various related matters, see
\cite{ah,pkt2,witten,eric1,hs}. The problem is further complicated in
the M-extended superspace, because of the presence of new coordinates.
To avoid the propagation of any unwanted degrees of freedom, one has to find 
new  kinds  of fermionic and bosonic local symmetries, analogous to the more 
familiar $\kappa$-symmetry.

\vspace{20pt}

\noindent{\bf Acknowledgements}

It is a pleasure to thank Eric Bergshoeff, Michael Duff and Yoshiaki
Tanii for useful discussions. I also would like to acknowledge the kind
hospitality extended to me at the International Center for Theoretical
Physics in Trieste, where this work was completed. 

\pagebreak

\end{document}